\documentclass[prl,twocolumn,nofootinbib]{revtex4}

\usepackage{amsmath}    
\usepackage{amsfonts}  
\usepackage{amssymb}
\usepackage{graphicx}   
\usepackage{dcolumn}
\usepackage{bm}
\usepackage{epsfig}

\usepackage{color}
\usepackage{hyperref}

\hypersetup{
    colorlinks=true,
    linkcolor=red,
    citecolor=blue,
}

\newcommand{\om}{\Omega_m}

\def\be{\begin{equation}}
\def\ee{\end{equation}}
\def\ba{\begin{eqnarray}}
\def\ea{\end{eqnarray}}

\begin{document}

\title{Testing Einstein Gravity with Cosmic Growth and Expansion}
\author{Gong-Bo Zhao$^{1,2}$, Hong Li$^{3,4}$, Eric V. Linder$^{5,6}$, Kazuya Koyama$^{2}$, David J. Bacon$^{2}$, Xinmin Zhang$^{3,4}$}
\affiliation{$^{1}$ National Astronomy Observatories, Chinese Academy of Science, Beijing, 100012, P.R.China}
\affiliation{$^{2}$ Institute of Cosmology \& Gravitation, University of Portsmouth, Portsmouth, PO1 3FX, UK}
\affiliation{$^{3}$ Theoretical Physics Division, Institute of High Energy Physics,
Chinese Academy of Science, P.O.Box 918-4, Beijing 100049, P.R.China}
\affiliation{$^{4}$ Theoretical Physics Center for Science Facilities (TPCSF),
Chinese Academy of Science, Beijing 100049, P.R.China}
\affiliation{$^{5}$ Berkeley Lab \& University of California, Berkeley, CA 94720, USA}
\affiliation{$^{6}$ Institute for the Early Universe WCU, Ewha Womans
University, Seoul, Korea}

\begin{abstract}

We test Einstein gravity using cosmological observations of both expansion
and structure growth, including the latest data from
supernovae (Union2.1), CMB (WMAP7), weak lensing
(CFHTLS) and peculiar velocity of galaxies (WiggleZ).  We fit modified
gravity parameters of the generalized Poisson equations simultaneously
with the effective equation of state for the background evolution,
exploring the covariances and model dependence.  The results show that
general relativity is a good fit to the combined data.  Using a 
Pad{\'e} approximant form for the gravity deviations accurately captures 
the time and scale dependence for theories like $f(R)$ and DGP gravity, 
and weights high and low redshift probes fairly.  For current observations, 
cosmic growth and expansion can be fit simultaneously 
with little degradation in accuracy, while removing the possibility of bias 
from holding one aspect fixed. 
\end{abstract}

\pacs{draft}

\keywords{large-scale structure, structure formation, modified gravity,
dark energy, theoretical cosmology}

\maketitle

The acceleration of the cosmic expansion was first discovered using the
supernovae Type Ia (SNIa) measurements in 1998 \cite{sn1998}, and later
confirmed by various independent cosmological probes including Cosmic
Microwave Background (CMB) \cite{wmap7}, Large Scale Structure (LSS)
\cite{lss}, Integrated Sachs-Wolfe (ISW) \cite{isw}, and so forth. A
crucial question is what is the physical origin of the acceleration.

Within the framework of general relativity (GR), a component in the
energy budget with substantially negative pressure, dubbed Dark Energy (DE),
is needed to drive the acceleration of the universe.  The
pressure to density ratio, or equation of state parameter, $w(z)$ is
an important characteristic to defining the physics responsible; for
example, quintessence behavior ($w>-1$) \cite{quintessence}, phantom
($w<-1$) \cite{phantom}, and quintom (where $w$ crosses $-1$ during
evolution) \cite{quintom} properties would each be an important clue.

Another approach to obtain the acceleration is to modify the gravity
theory. In this scenario, no dark energy component exists, but the laws of
gravity relating spacetime curvature to the material contents are
changed in such a way as to drive acceleration; the modified terms can be
viewed as an effective dark energy contribution with some effective
equation of state.  

In terms of only the background expansion, these two approaches are
indistinguishable.   As well, the growth of structure is suppressed by the
acceleration and is further affected by modifying the gravitational laws.
These can be offset in such a way that a modified gravity (MG) model and
some dark energy (DE) model have the same growth evolution.  However, the
key point is that they will not in general simultaneously have the same
expansion {\it and\/} same growth behavior.  

This highlights the importance of simultaneously fitting the data both for 
possible gravitational modifications and the expansion history.  However,
while much effort in the literature has gone into comparing MG models with
the observational data, in almost all cases a $\Lambda$CDM background is
assumed.  This can lead to bias in the gravity parameters derived (if the
true cosmology does not have the $\Lambda$CDM expansion) by a statistically
significant amount, even if $\langle w\rangle=-1$ \cite{hutlin}.  Moreover,
this can cause further errors since $w$ can be strongly correlated with
other cosmological parameters such as neutrino mass, spatial curvature,
the tensor perturbations, etc.\ \cite{Zhao:2006qg,Xia:2008ex,Li:2008vf}.

In this paper, we aim to test GR using the latest observations, including
SNIa, CMB, weak lensing (WL), and the peculiar velocity field of galaxies
(PV), while specifically fitting the background cosmology simultaneously
with the MG parameters.  While simultaneous fitting has been considered
before, for future data projections, e.g.\ \cite{stril}, we use actual
data, employ a more comprehensive parametrization of gravity
modifications, and study the covariance  between MG parameters and the
background equation of state in more detail.

In Newtonian gauge, the metric in a perturbed FRW universe reads,
\be\label{eq:metric}
ds^2=-(1+2\Psi)dt^2+(1-2\Phi)a^2\delta_{ij}dx^idx^j\,,
\ee
where $\Phi$ and $\Psi$ denote the space curvature perturbation and the
gravitational potential respectively, and they are related to the comoving
matter density perturbation $\Delta$ via ~\cite{Song:2010fg,Pogosian:2010tj}
\ba
\label{eq:MG_para1}k^2\Psi &=& -4\pi G a^2\mu(k, a) \rho \Delta, \\
\label{eq:MG_para2}k^2 (\Phi + \Psi) &=& -8\pi G a^2 \Sigma(k, a) \rho \Delta,
\ea
where $G$ is Newton's gravitational constant, $\rho$ is the homogeneous
matter density, $k$ is the wavenumber, and $a$ is the expansion scale factor.

The functions $\mu$ and $\Sigma$ are both unity in GR, but in general they
can be functions of both scale and time in modified gravity. The quantity
$\mu$ in the Poisson equation, Eq.~(\ref{eq:MG_para1}),  determines the
modified growth with respect to that in GR, which can be measured using
the peculiar velocities of galaxies (or their density field, but this
involves a factor of the generally unknown bias factor relating mass to
galaxy light).  The quantity $\Sigma$ can be constrained by the weak
lensing measurement since it is directly related to the lensing potential
$\Phi + \Psi$.  Indeed one can view the first equation as governing motion
along geodesics of nonrelativistic tracers while the second involves the
null geodesics of  light.  Therefore, the PV and WL measurements are highly
complementary to probe for the deviation from GR encoded in the functions $\mu$
and $\Sigma$ \cite{Song:2010fg}.  Note that in \cite{Daniel:2010yt} the
variable $\mu$ is called $\mathcal{V}$ and $\Sigma={\mathcal G}$, and a
translation table is provided there for other functions in the literature.

One still needs to parametrize the two functions of wavenumber and scale
factor.  There are many possibilities, including principal component
analysis \cite{Zhao:2009fn}, bins \cite{Daniel:2010yt}, etc.
In this paper we propose a parametrisation for $\mu$ and $\Sigma$ that
covers a wide range of modified gravity models known so far. This
parametrisation is based on Brans-Dicke, or more generally scalar-tensor,
gravity. Using the quasi-static approximations, the
solutions for $\mu$ and $\Sigma$ are obtained as
\begin{eqnarray}
\mu &=& \frac{G(a)}{G_N} \left(1+ \frac{1}{3\frac{a^2 M(a)^2}{k^2} + 2 \omega_{BD}(a)+3}
\right), \\
\Sigma &=& \frac{G(a)}{G_N},
\end{eqnarray}
where $G$ is the gravitational coupling generalizing Newton's constant
$G_N$ measured locally by the Cavendish-type experiments, $M$ is the mass
of the scalar field and $\omega_{BD}$ is the Brans-Dicke (BD) parameter.
Under the quasi-static approximations, $G, M$ and $\omega_{BD}$ can weakly
depend on time, i.e.\
$\dot{G}/G, \dot{M}/M, \dot{\omega}_{BD}/\omega_{BD} \ll k/a$. There are
two possibilities to recover GR, i.e. $\mu \to 1$, in this Ansatz.
One is to have the Compton wavelength of the scalar field smaller than
the scales of interest, $k/(a M) \ll 1$, and the other is to consider
a large BD parameter $\omega_{BD} \gg 1$.

Since the time variation of the generalized Newton's constant is strongly
constrained, we take $G(a)=G_N$ and consider the following two cases for
simplicity: (i) the Brans-Dicke parameter vanishes $\omega_{BD} =0$, or 
(ii) the scalar field is massless $M=0$.  Essentially we are considering
the sources of modification one by one, and in these cases $\mu$ and
$\Sigma$ can be parametrised in a very simple form: 
\begin{eqnarray}
\mu=1+\frac{c a^s k_H^n}{1+3c a^s k_H^n} \,, \quad
\Sigma=1 \,. \label{eq:muc}
\end{eqnarray}
Here, in case (i) we have $c=(3M_0^2/H_0^2)^{-1}$, $n=2$, and $k_H$ denotes 
the dimensionless wavenumber, namely, $k_H\equiv k/H_0$ where $k$ and $H_0$ 
are the usual wavenumber and the Hubble constant, respectively; in
case (ii) we have $c=1/(2\omega_{BD,0})$ and $n=0$; thus $c$ quantifies
the dimensionless amplitude of deviations from GR, with $c\ll 1$ recovering 
GR, and has the physical interpretation in terms of either the ratio today 
of the scalar Compton wavelength to the Hubble scale, or the BD parameter.
The time variation of either $(aM)^2$ or $\omega_{BD}$ is approximated as a
power law $a^{-s}$ (so GR is recovered in the early universe for $s>0$),
and the spatial variation is set to either $k^2$ or scale independence.
These are motivated by known theories of modified gravity as we discuss below.

Case (i) includes $f(R)$ gravity.  For $f(R)$ theory, the deviations depend
on the scalaron mass, $M(a)=1/\sqrt{3d^2f/dR^2}$, which defines a Compton
length over which the deviations propagate.  In general, in the early
universe or in high curvature regions, $M\gg H$ and the equations reduce
to general relativity.  The parametrization $M(a)=M_0 a^{-\sigma}$ has been
shown to be accurate over the past evolution by \cite{bz08,Zhao:2008bn,aw10}.
The deviation of the MG parameter $\mu(k,a)$ from unity within $f(R)$ theory
is \cite{Zhao:2008bn}
\begin{equation}
\mu(k,a)=1 + \frac{1}{3+3(aM/k)^2} \,.
\end{equation}
Thus within this ansatz for $f(R)$, this leads to Eq.~(\ref{eq:muc}) with
$s=2(\sigma-1)$.

Case (ii) includes DGP gravity \cite{Dvali:2000hr}.  This has no scale
dependence (on cosmological scales, much greater than the Vainshtein scale).
For DGP gravity the deviation is given by
\begin{equation}
\mu(k,a)=1-\frac{1}{3} \,\frac{1-\om^2(a)}{1+\om^2(a)} \,,
\end{equation}
where $\om(a)=\om a^{-3}/(H/H_0)^2$ is the dimensionless matter density
as a function of scale factor, with $\om$ the value today.
This corresponds to $2 \omega_{BD}(a) = -6/[1-\om^2(a)]$.  Note that
in the matter dominated epoch when $\om(a)\to1$ then GR is recovered,
while in the future when the matter density redshifts away then
$\omega_{BD}\to -3$.  In general $\omega_{BD}$ does not behave
as a power law in scale factor: at high redshift $s=3/2$, then it steepens,
before evolving toward $s=0$ in the asymptotic future.  However, a
reasonable fit to the function $\mu$ can be achieved by the Pad{\'e}
approximant form of Eq.~(\ref{eq:muc}).



For the background expansion, we include an effective dark energy equation
of state (EOS)  $w(a) = w_0+w_a(1-a)$
known to be highly accurate in describing the background expansion for a
wide variety of models \cite{lin2003,depl2008}.

Our full parameter set to be constrained using the current observational
data consists of the MG parameters $c, s$, the expansion parameters
$w_0,w_a$ and the other cosmological parameters.  Specifically,
we parametrise the universe using
\be
\label{eq:paratriz} {\bf P} \equiv (\omega_{b}, \omega_{c},
\Theta_{s}, \tau, n_s, A_s, c, s, w_0, w_a,\mathcal{N}),
\ee
where
$\omega_{b}\equiv\Omega_{b}h^{2}$ and
$\omega_{c}\equiv\Omega_{c}h^{2}$ are respectively the physical baryon and
cold dark matter densities relative to the critical density,
$\Theta_{s}$ is the ratio (multiplied by 100) of the sound
horizon to the angular diameter distance at decoupling, $\tau$
denotes the optical depth to re-ionization, and $n_s$ and $A_s$ are the
primordial density power spectral index and amplitude respectively.
We also vary, and marginalize over, several astrophysical nuisance
parameters denoted by $\mathcal{N}$ when performing our likelihood
analysis, including those associated with the galaxy distribution for
WL data and the absolute luminosity for supernovae.

The datasets used are of weak lensing (the two point correlation function
$\xi_E$ at $\theta>30'$ from the CFHTLS-Wide survey \cite{Fu:2007qq, Kilbinger:2008gk}, the same dataset used in Ref. \cite{Song:2010fg}),
peculiar velocities ($f\sigma_8$ in four redshift bins in the range of $z\in[0.1,0.9]$ measured by the WiggleZ team using the redshift space distortion measurement
\cite{Blake:2011ep})\footnote{We 
did not use peculiar velocity data when constraining the $n=2$ 
models since scale independent growth had already been assumed 
in extracting the data.}, 
supernovae distances from the Union2.1 compilation 
including systematic errors \cite{Suzuki:2011hu}, the full CMB spectra of
WMAP seven year data \cite{wmap7}, and baryon acoustic oscillation distance
ratios from SDSS DR7 galaxies \cite{lss}. We do not directly employ the galaxy
density power spectrum so as to avoid uncertainties in galaxy bias, which
in principle could give time- and scale-dependent signatures similar to
modified gravity.

Given the set of cosmological parameters ${\bf P}$ in
Eq.~(\ref{eq:paratriz}), we calculate the theoretically expected
observables (CMB spectra, luminosity distance, velocity growth factor
$\Theta$, and the E-mode component $\xi_E$ of the weak lensing shear) 
using {\tt MGCAMB} \cite{Zhao:2008bn}.  We then constrain the model
parameters using a version of the Markov Chain Monte Carlo (MCMC) package
{\tt CosmoMC}~\cite{Cosmomc, Lewis:2002ah} modified to include our extra
parameters.  We impose priors on $c, s$ of $c\ge-1/3$ for $n=0$ (to avoid 
the pole in $\mu$, corresponding to excluding $-3/2 < \omega_{BD,0} <0$), 
and $c\ge0$ for $n=2$, and $s\in[1,4]$.


The results are summarised in Table~\ref{table1}.  We run three different
types of models: a scale independent case (``DGP''), a scale dependent
$k^2$ case (``$f(R)$''), and a true dark energy case fixing $c=0$ and
including dark energy perturbations in the calculations \cite{Zhao:2005vj}.  
The time dependence parameter $s$ cannot be constrained by the data and 
is marginalized over.

\begin{table}[t]
\begin{tabular}{c|c|c|c|c|c|}
   \cline{2-6}
    & \multicolumn{2}{c|}{scale indep. ($n=0$)} &
\multicolumn{2}{c|}{scale dep. ($n=2$)} &\multicolumn{1}{c|}{GR: $\mu=1$} \\  \cline{2-6}
    & $w=-1$ & $w_0,w_a$ float& $w=-1$ & $w_0,w_a$ float & $w_0,w_a$ float\\ \hline
    \multicolumn{1}{|c|} {$c$} &
$<4.0$  &$<4.1$ & $<0.002$& $<0.002$ &$0$\\  \hline
    \multicolumn{1}{|c|} {$w_0$} &
$-1$ & $-0.90\pm0.19$& $-1$& $-0.92\pm0.20$ &$-0.91\pm0.19$\\  \hline
    \multicolumn{1}{|c|} {$w_a$} &
$0$  & $-0.26\pm0.78$& $0$& $-0.32\pm0.82$ &$-0.27\pm0.78$\\  \hline

\end{tabular}
\caption{Constraints from current data on the MG parameter $c$ (marginalized
over $s$) and the effective dark energy parameters $w_0, w_a$ for the scale 
independent, scale dependent MG models and the true dark energy model 
($\mu=1$). For $c$ we quote the 95\% CL upper 
limit, while for $w_0$ and $w_a$, we quote the median and 68\% CL error. 
Fitting for the expansion in terms of $w_0,w_a$ rather than fixing $w=-1$ 
does not degrade the gravity constraint; fitting for gravity in terms of 
$c,s$ does not degrade the expansion constraint. 
}
\label{table1}
\end{table}

\begin{figure}[t]
\includegraphics[scale=0.157]{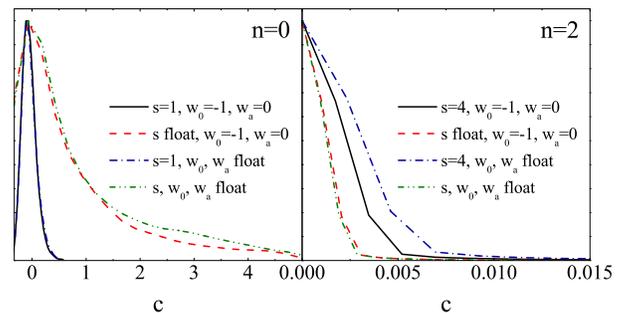}
\caption{1D PDF for the MG amplitude $c$ from the latest observational
data.  The left panel shows the scale independent case, with $s$ either
fixed to 1 or marginalized over, and the background either fixed to
$\Lambda$CDM or marginalized over $w_0,w_a$. 
The right panel shows the analogous curves for the scale dependent ($k^2$)
case. 
All PDFs are consistent with $c=0$, corresponding to GR. }
\label{fig:c1D}
\end{figure}

Figure~\ref{fig:c1D} shows the 1D probability distribution functions
(PDF) for $c$.  Recall that $c=0$ is GR, and we see that all cases are
consistent with GR.  
For each case we either also fit $s$ or fix it to $s=1$ for the 
scale independent case (mimicking DGP) or to $s=4$ for the $k^2$ case
(mimicking a particular $f(R)$).  
Note that when we marginalize over $s$ this can actually tighten the 
constraints on $c$ because small values of $s$ are then permitted (recall 
deviations depend on $a^s$), which strengthens deviations at higher
redshifts. 

\begin{figure}[t]
\includegraphics[scale=0.155]{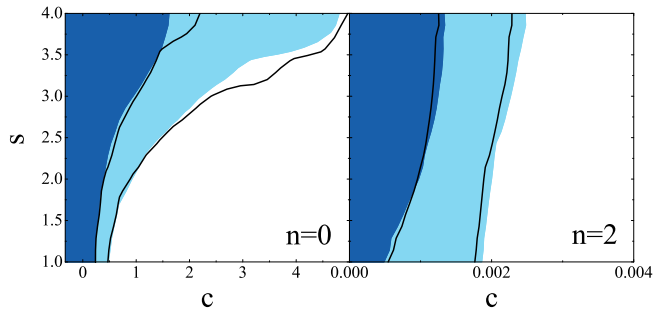}
\caption{68\%  and 95\% CL joint constraints on the MG parameters.
Filled and unfilled contours represent the cases when the background
cosmology is fixed to $\Lambda$CDM, and the effective dark energy equation
of state parameters $w_0, w_a$ are allowed to vary, respectively.  GR
corresponds to $c=0$, when the value of $s$ is moot.  The left (right) panel
corresponds to the scale independent (dependent, $k^2$) case.
}
\label{fig:cs2D}
\end{figure}

Figure~\ref{fig:cs2D} contains the 2D joint probability
contours for $c$--$s$ for scale dependent and independent cases.  The filled
contours represent when the background expansion is fixed to $\Lambda$CDM;
we see that this does not have a dramatic effect on the results, implying
that there is little covariance between the gravity and expansion
parameters and that simultaneous fitting is not only desirable but practical.

\begin{figure}[t]
\includegraphics[scale=0.157]{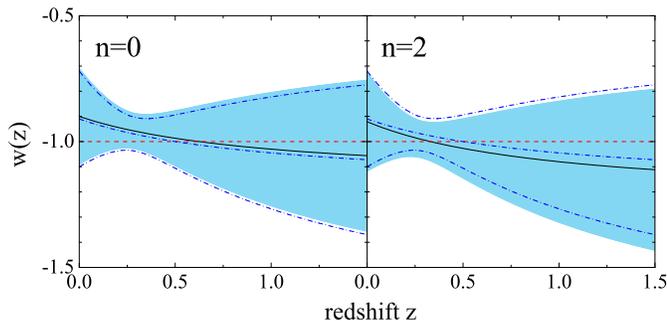}
\caption{The reconstructed $w(z)$ with 68\% CL error are shown allowing
for modified gravity (marginalized over $c, s$) in the scale independent 
(left panel) and scale-dependent $k^2$ (right panel) cases by the filled
bands.  The reconstruction for true dark energy, with gravity fixed to GR,
is shown by the dash-dotted curves, the same in each panel. }
\label{fig:w0wa}
\end{figure}

In Fig~\ref{fig:w0wa}, we show the reconstructed (effective) $w(z)$ using 
the constraints on the expansion parameters for both MG models and for the 
true dark energy case, with $c$ and $s$ marginalized over where 
appropriate.  The consistency of the contours demonstrates that simultaneous 
fitting of gravity and expansion does not here substantially degrade 
constraints.  Recall that the simultaneous fitting 
enables avoidance of a possible significant bias if there is any deviation 
from $\Lambda$CDM with GR.  Current data is consistent with $\Lambda$CDM 
cosmology even in the presence of possible gravitational modifications.

The lack of covariance between the gravity and expansion parameters is
not a general property true for all large scale structure observations;
for example \cite{stril} found high correlation between the gravitational
growth index $\gamma$ \cite{groexp} (closely related to $\mu$) and
$w_0, w_a$ when using the galaxy
density power spectrum for future data.  This is because the density
power spectrum involves the integrated growth factor, influenced by
both expansion and  modified gravity, while the PV field involves the
growth rate currently at modest accuracy and current weak lensing probes 
mostly light deflection.
With future data, however, weak lensing will be more sensitive to growth,
and galaxy data will probe both growth and growth rate, so we expect
that as the constraints tighten they will also become more correlated.
Of course both the EOS and MG parameters have covariance
with the matter density $\Omega_m$, thus in principle they are correlated
indirectly. But for our current datasets the correlation is seen to be
very small.

Indeed the constraints on $w(z)=w_0+w_a(1-a)$ are nearly independent of
the MG parameters.  A similar behavior was shown in Fig.~2 of
\cite{linroysoc} where the joint confidence contours of $w_0$--$w_a$
were nearly independent of the value of the modified growth index $\gamma$.
The $w(z)$ behavior reconstructed from our best fit shows the usual
``quintom'' \cite{quintom} crossing of $-1$ at $z\sim0.4$.  Such a 
pivot point is
expected from the strong influence of CMB constraints on the distance to
last scattering agreeing with $\Lambda$CDM; this induces the ``mirage of
$\Lambda$'' \cite{linmirage} where $w_p\equiv w(z\approx0.4)\approx -1$
even in the presence of time variation in EOS, and is not a consequence
of using the $w_0$, $w_a$ form.  As other data gain in leverage relative
to this CMB geometric constraint, the crossing may disappear.  On the
other hand, the effective EOS in modified gravity models can cross $-1$,
so this quintom behaviour, if confirmed by future data to high confidence
level, might be a smoking gun of modified gravity.



In summary, to test gravity in a stringent manner, we parametrize the consistent set
of gravity field equations through modifying factors in the matter growth 
(Poisson) equation and light
deflection (sum of the metric potentials) equation.  The Pad{\'e}
approximant form adopted can cover many different modified gravity theories
with scalar degrees of freedom.  Note that a simple power law such as
$\mu=1+\mu_s a^s$ cannot properly weight both high and low redshifts and
may bias the results.
Simultaneously with fitting for these modifications we
also allow the background expansion to deviate from a $\Lambda$CDM cosmology
through an effective time varying dark energy equation of state.  This is 
important 
as incorrectly fixing either the gravity side
or the expansion side could strongly bias the conclusions.

We then used the most recent observational data -- supernova distance data
(Union2.1 compilation), CMB (full WMAP-7yr spectra), weak lensing (CFHTLS),
and galaxy peculiar velocity (WiggleZ) -- to fit these and other
cosmological parameters, testing Einstein gravity.  The simultaneous
fitting of gravity and expansion can be successfully carried out, with
little degradation in leverage while avoiding possible bias due to fixing
one or the other. 
In the scale dependent case, the deviation amplitude is constrained to be 
$c\lesssim0.002$ corresponding to the constraint on the Compton wavelength 
$\lambda\lesssim250\,h^{-1}$Mpc (95\% CL), while in the scale independent case 
cosmological data is not yet precise enough to place strong bounds 
($c\lesssim4.1$ implies $\omega_{BD,0}\gtrsim0.12$). 

General relativity is a good fit with these recent data.  Future data
will allow more stringent limits, and as growth measurements improve the
covariance between gravity and expansion influences should increase,
making simultaneous fitting even more necessary.  The function $\Sigma$
entering the light deflection equation will be tested as upcoming large
weak lensing surveys deliver data.  Next generation data should greatly
advance our ability to test gravity and uncover the physical origin of
the acceleration of our universe.

We thank the Supernova Cosmology Project for providing the Union2.1 data 
before publication.  GZ and KK are supported by STFC
grant ST/H002774/1; EL is supported by DOE and by WCU grant R32-2009-000-10130-0. KK is also supported by the ERC and the Leverhulme
trust. DB acknowledges the support of an RCUK Academic Fellowship. HL and XZ are supported in part by the National Natural Science
Foundation of China under Grant Nos. 11033005, 10803001, 10975142
and also the 973 program No. 2010CB833000.

\end{document}